\documentclass[prc,twocolumn,floatfix,%
showpacs,preprintnumbers,amsmath,amssymb]{revtex4}
\usepackage{graphicx}
\def\drnp{\Delta r_{np}} \def\apv{A_{pv}} \def\pbx{$^{208}$Pb }

\begin{document}
\title{Neutron skin of $^{208}$Pb, nuclear symmetry energy,
and the parity radius experiment}
\author{X. Roca-Maza$^{1,2}$} \author{M. Centelles$^1$} 
\author{X. Vi\~nas$^1$} \author{M. Warda$^3$}
\affiliation{$^1$Departament d'Estructura i Constituents de la Mat\`eria
and Institut de Ci\`encies del Cosmos, Facultat de F\'{\i}sica, Universitat de Barcelona,
Diagonal {\sl 647}, {\sl 08028} Barcelona, Spain\\
$^2$INFN, sezione di Milano, via Celoria 16, I-{\sl 20133} Milano, Italy\\ 
$^3$Katedra Fizyki Teoretycznej, Uniwersytet Marii Curie--Sk\l
odowskiej, ul.\ Radziszewskiego {\sl 10}, {\sl 20-031} Lublin, Poland}

\begin{abstract}
A precise determination of the neutron skin $\drnp$ of a heavy nucleus
sets a basic constraint on the nuclear symmetry energy ($\drnp$ is the
difference of the neutron and proton rms radii of the nucleus). The parity
radius experiment (PREX) may achieve it by electroweak parity-violating
electron scattering (PVES) on $^{208}$Pb. We investigate PVES in nuclear
mean field approach to allow the accurate extraction of $\drnp$ of \pbx
from the parity-violating asymmetry $\apv$ probed in the experiment. We
demonstrate a high linear correlation between $\apv$ and $\drnp$ in
successful mean field forces as the best means to constrain the neutron
skin of \pbx from PREX, without assumptions on the neutron density shape.
Continuation of the experiment with higher precision in $\apv$ is
motivated since the present method can support it to constrain the density
slope of the nuclear symmetry energy to new accuracy.\end{abstract}
\date{\today}\maketitle

New interest in masses and density distributions of nuclei is being
prompted by the production of rare isotopes in radioactive beam facilities
\cite{enam08}. Exciting phenomena discovered in these isotopes such as thick
skins, halos, and new shell closures urge better understanding of neutrons
in nuclei. Yet, our knowledge of neutron density distributions is limited
even in the stable nuclei. As neutrons are uncharged, neutron densities
have been probed mostly by nucleon scattering \cite{hof80,zen10}, $\alpha$
scattering \cite{kra04}, and nuclear effects in exotic atoms
\cite{klo07,fried09}. Even if some of these experiments reach small
errors, all hadronic probes require model assumptions to deal with the
strong force introducing possible systematic uncertainties.

Parity-violating electron scattering (PVES) was suggested as a {\em
model-independent} probe of neutron densities \cite{don89}. An electroweak
probe is not hindered by the complexity of the strong force and the
reaction mechanism with the nucleus needs not be modeled \cite{don89,vret00,prex2},
similarly to clean electron scattering for nuclear charge densities.
The novel parity radius experiment (PREX) at the Jefferson Lab
\cite{prex1,prex2} aims to measure the parity-violating asymmetry $\apv$
in polarized electron scattering on \pbx to 3\% accuracy. This
accuracy is estimated to constrain the neutron rms radius $r_n$ of
\pbx to 1\% \cite{prex1,prex2}. Currently, $r_n$ of \pbx is
uncertain by $\sim$2\% and data may be model dependent
\cite{hof80,zen10,kra04,klo07,fried09,cen10}; in contrast, the charge 
radius of \pbx is accurately known as $r_{\rm ch} = 5.5010(9)$ fm
\cite{ang04}. In recent years it has been established that the neutron
skin thickness $\drnp = r_n - r_p$ (difference of the neutron and
proton rms radii) of \pbx is strongly {\em correlated with} the
density dependence of the nuclear symmetry energy around saturation
\cite{bro00,fur02,ste05a,todd05,cen09}. Knowledge of the density
dependence of the nuclear symmetry energy is a cornerstone for drip lines,
masses, densities, and collective excitations of neutron-rich nuclei
\cite{bro00,fur02,ste05a,cen09,todd05,chen10,carbone10},  flows and
multifragmentation in heavy-ion collisions \cite{li08,tsa09}, and for
astrophysical phenomena like supernovae, neutrino emission, and neutron
stars \cite{hor01,ste05a,xu09,stei10}. A constraint from PREX on $\drnp$
of \pbx is thus regarded as a landmark for isospin physics.
In addition to being important for its own sake, it has broad implications for
different communities of nuclear physics and astrophysics. Fostered by
the seminal study of Ref.\ \cite{prex2}, PREX completed an initial run in
2010. First analyses \cite{prex1} show the validity of the experimental
technique, the adequacy of instruments, and that systematic errors are
under control. Additional beam time is now under request to attain the
planned 3\% accuracy in the parity-violating asymmetry $\apv$ \cite{prex1}.

The direct output of PREX is the value of the asymmetry $\apv$ at a
single scattering angle \cite{prex1,prex2}. The neutron rms radius $r_n$
of the nucleus may be deduced only if a shape for the neutron density such
as a two-parameter Fermi function \cite{prex2} is assumed.~A systematic
uncertainty in the analysis is unavoidable in this way. Here, we provide a
different and accurate strategy to deduce $r_n$ and $\drnp$ from
PREX that removes this problem. By study of PVES on \pbx in
successful nuclear mean field (MF) forces of wide use in nuclear research
and astrophysical applications, we reveal a high linear relation between
$\drnp$ and $\apv$ that allows one to extract $r_n$ and $\drnp$
from $\apv$ model and shape independently. Moreover, our
approach unifies the extraction of $\drnp$ from $\apv$ with the
same framework where $\drnp$ is correlated to the symmetry energy.
We show that the present method can support PREX to narrow down the value
of the density slope of the nuclear symmetry energy to novel accuracy.
This result provides a new and important motivation to continue the
experiment to increased precision.

Electrons interact with nuclei by exchanging photons and $Z^0$ bosons. The
former mainly couple to protons and the latter to neutrons because,
opposite to the nucleon electric charges, the neutron weak charge
$Q^n_{\rm W}=-1$ is much larger than the proton weak charge $Q^p_{\rm
W}=1-4\sin^2\theta_{\rm W}\approx0.075$ ($\theta_{\rm W}$ being the
Weinberg angle). Therefore, electron scattering can probe both the
electric and the weak charge distributions in a nucleus
\cite{don89,prex2,vret00}. PREX measures the elastic differential cross
sections $d\sigma_\pm/d\Omega$ for incident electrons of positive or
negative helicity. The parity-violating asymmetry,
\begin{equation}\apv=
\Big(\frac{d\sigma_+}{d\Omega}-\frac{d\sigma_-}{d\Omega}\Big) \bigg/
\Big(\frac{d\sigma_+}{d\Omega}+\frac{d\sigma_-}{d\Omega}\Big) 
\label{apv}\end{equation}
for massless electrons (it is $m_e/p_e\approx 0.0005$ at PREX energy), is
sensitive to the parity-violating term induced by the weak interaction in
the scattering amplitude. According to their helicity, electrons interact with
a potential $V_{\rm Coulomb }(r) \pm G_{F} \rho_{\rm W}(r) /2^{3/2}$, with
$G_{F}$ the Fermi constant and $\rho_{\rm W}$ the weak density of the target
\cite{don89,vret00,prex2}. We solve the associated Dirac equation via the
exact phase-shift analysis in distorted-wave Born approximation (DWBA)
\cite{cen10} to compute $\apv$. Our benchmarks are the pointlike densities
of protons $\rho_p(r)$ and neutrons $\rho_n(r)$ calculated
self-consistently in MF models. We fold $\rho_p(r)$ and $\rho_n(r)$ with
electromagnetic proton and neutron form factors to obtain the charge
density \cite{cen10}, and with electric form factors for the coupling to a
$Z^0$ to obtain the weak density \cite{cen10,prex2,moya10}:
$\rho_{\rm W}(r)= \int\! d{\bf r}'\,\{\,
4\,G_n^{Z^0}\!(r')\,N\rho_n(|{\bf r} -{\bf r}'|)+
4\,G_p^{Z^0}\!(r')\,Z\rho_p(|{\bf r} -{\bf r}'|)\,\}$.

Though not useful for realistic calculations, it is worth recalling
the Born approximation (BA) to $\apv$ \cite{don89,prex2}:
\begin{equation} \apv^{\rm BA}= \frac{G_F q^2}{4\pi \alpha \sqrt{2}}
 \Big[ 4 \sin^2\theta_{\rm W} - 1 + \frac{F_n(q)}{F_p(q)} \Big],
\label{apvpwba}\end{equation}
as it nicely illustrates that $\apv$ relates to the neutron and
proton nuclear form factors $F_{n,p}(q)$. Furnstahl \cite{fur02} showed
that $F_n(q) = (4\pi)^{-1}\! \int\!d^3 r\, j_0(qr) \rho_n(r)$ is at low
momentum transfer $q$ strongly correlated with $r_n$ of \pbx in
nuclear MF models, evidencing that PREX would directly constrain the
neutron radius and the symmetry energy. Realistic DWBA calculations of
$\apv$ in MF models can be found in \cite{vret00,moya10,cen10,ban10}.

\begin{figure}[t]
\includegraphics[width=0.95\columnwidth,clip=true]{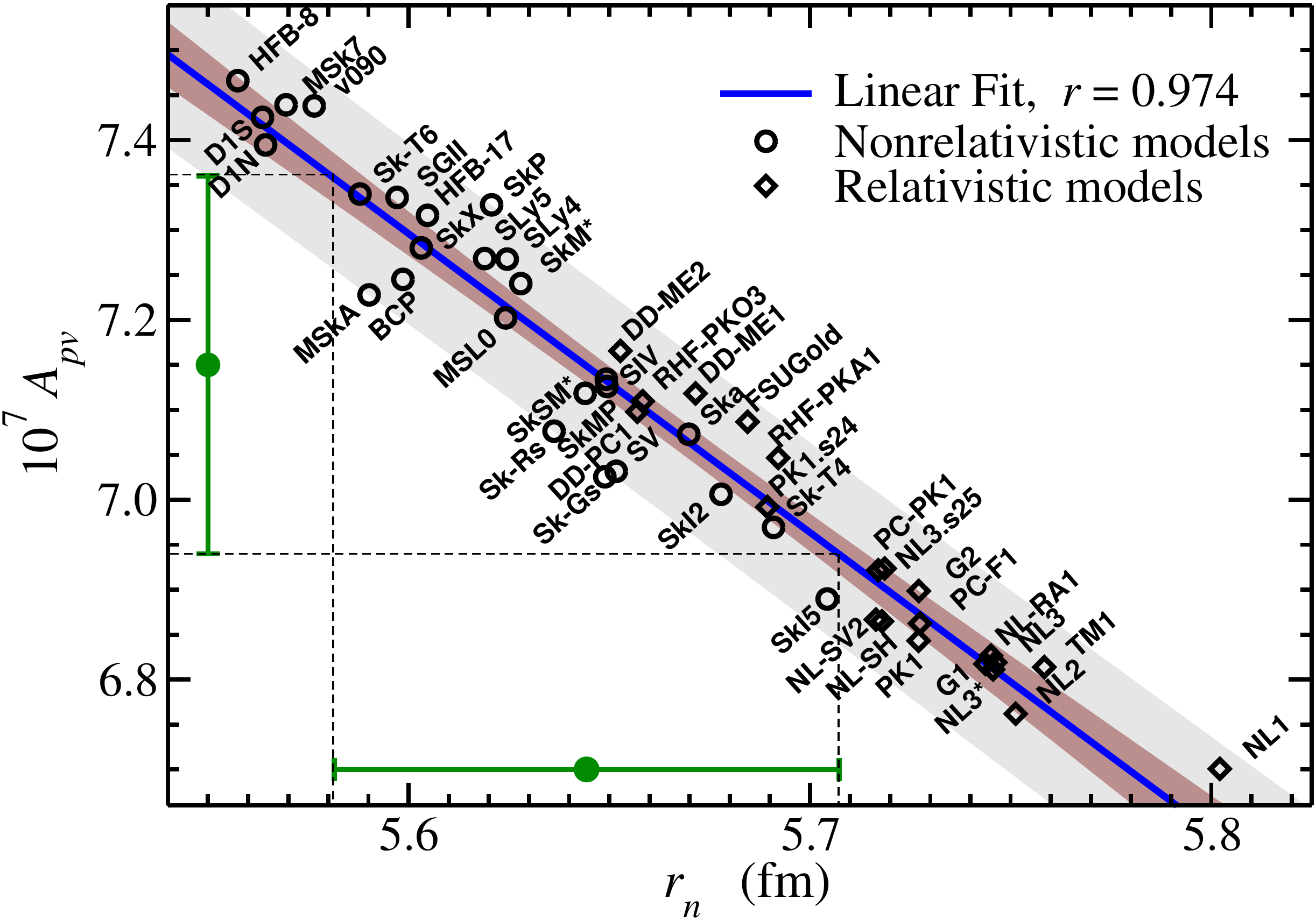}
\caption{\label{alrrn}
Parity-violating asymmetry for \pbx at the kinematics of PREX
against the neutron radius of \pbx in nuclear models. The linear
fit is $10^7 \apv= 25.83 - 3.31 r_n$. The inner/outer colored regions
depict the loci of the 95\% confidence/prediction bands of the regression
(see e.g.\ Ch.\ 3 of \cite{regres}). An assumed sample measurement
$\apv=0.715$ ppm of 3\% accuracy and its projection on the $r_n$ axis are
also drawn.}\end{figure}

At the optimal kinematics of PREX the electron beam energy is 1.06 GeV and
the scattering angle is $5^\circ$ ($q_{\rm lab}\!\approx\! 0.47$
fm$^{-1}$) \cite{prex1}. We compute $\apv$ in DWBA at this kinematics in
a comprehensive large sample of 47 nuclear MF interactions. We display the
results in Fig.\ \ref{alrrn} as a function of the neutron rms radius of
$^{208}$Pb. To prevent eventual biases in our study, we avoid including
more than two models of the same kind fitted by the same authors and
protocol. We also avoid models yielding a charge radius of \pbx away
from experiment \cite{ang04} by more than 1\% (same level as the 1\%
pursued by PREX in $r_n$). The considered models rest on very different
theoretical grounds, from {\em nonrelativistic} models of zero range
(models HFB, v090, and those starting with S or M) or finite range (D1S,
D1N, BCP), to {\em relativistic} models with meson self-interactions (NL
and PK models, FSUGold, G1, G2, TM1), density-dependent vertices (DD-ME,
RHF-PK) or point couplings (DD-PC1, PC-PK1, PC-PF1)
\cite{cen09,cen10,carbone10,chen10}. (NL3.s25 and PK1.s24 are variants of NL3 and
PK1 giving $\drnp=0.25$ and 0.24 fm in $^{208}$Pb.) All such models
accurately describe general properties of nuclei such as binding energies
and charge radii along the periodic table. However, one readily sees in
Fig.~\ref{alrrn} that the predicted $r_n$ of \pbx varies largely, from
5.55 to 5.8 fm, as the isovector channel of the nuclear models is little
constrained by current phenomenology. The models with softer (stiffer)
symmetry energy at saturation density \cite{cen10} yield smaller (larger)
$r_n$ and larger (smaller) $\apv$. One notes that the information encoded
in the models implies a value of about 0.67 to 0.75 ppm for $\apv$ at PREX
kinematics. A significant linear trend is found between $\apv$ and $r_n$
(the correlation coefficient is $r=0.974$).

As the experimental value of $\apv$ is not yet available, we have chosen
for study a plausible {\em test} value $\apv=0.715$ ppm {\em of\/ {\rm
3}\% accuracy}, depicted in Fig.~\ref{alrrn}. The assumed sample
measurement of $\apv$ determines through the linear fit shown in
Fig.~\ref{alrrn} a fiducial neutron rms radius $r_n=5.644\pm0.065$ fm,
within typical values deduced from hadronic probes
\cite{hof80,zen10,kra04,klo07,fried09}. Note that a 3\% accuracy in
$\apv$ does lead to $\sim$1\% accuracy in $r_n$, thereby supporting the
expectations of PREX. It is to be pointed out that the analysis described
in this paper is actually independent of the exact value of the
parity-violating asymmetry. Thus, once the experimental value of PREX is
known, one can repeat the same type of analysis using the actual $\apv$
instead of our test value. We also plot in Fig.~\ref{alrrn} the confidence
band of the regression (boundary of the possible straight lines) and the
so-called prediction band (the wider band that basically coincides with
the envelope of the models in the figure) at 95\% confidence level \cite{regres}.

\begin{figure}[t]
\includegraphics[width=0.95\columnwidth,clip=true]{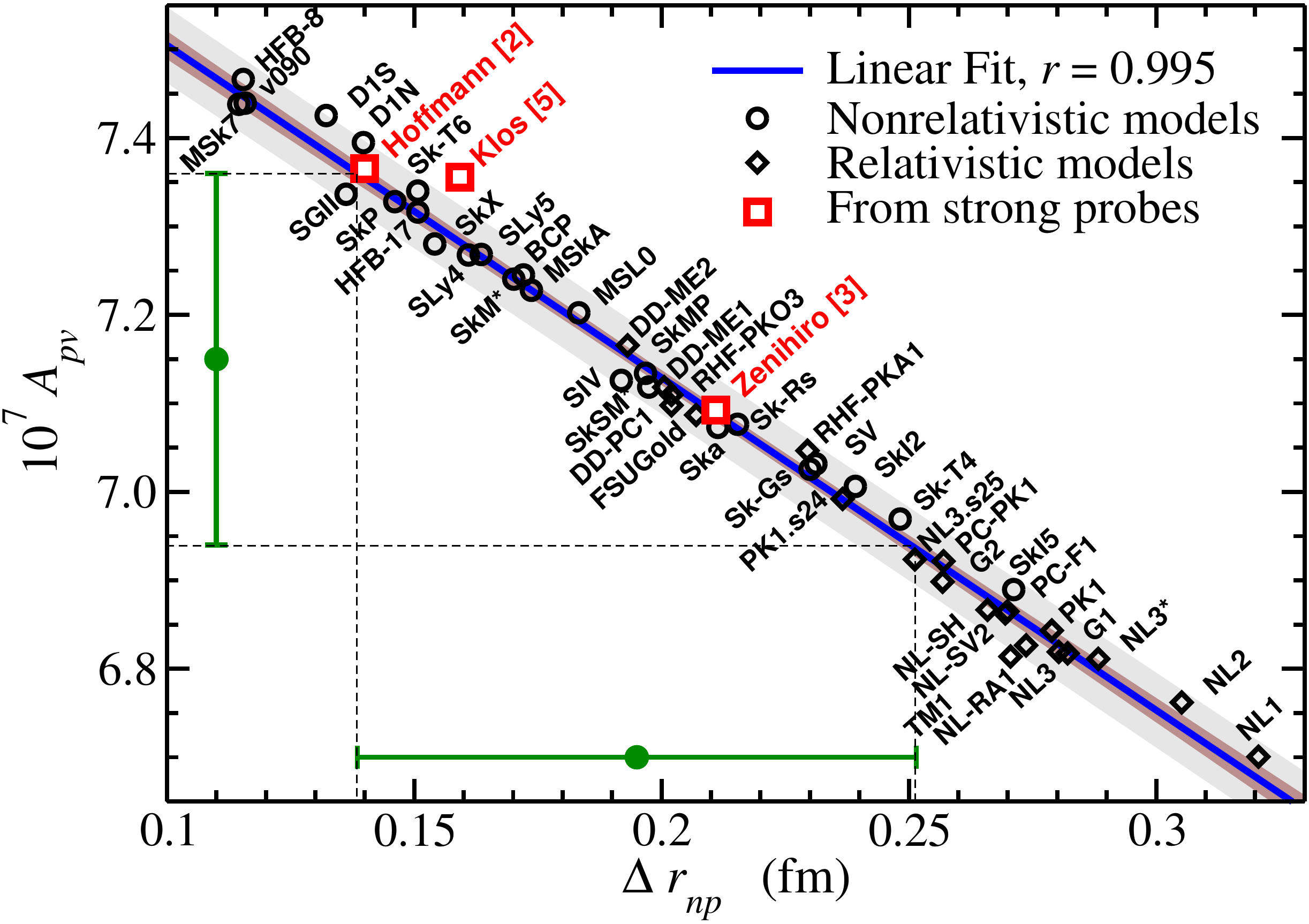}
\caption{\label{alrrnp}
Same as Fig.~\ref{alrrn} against the neutron skin of $^{208}$Pb. The
linear fit is $10^7 \apv= 7.88 - 3.75 \drnp$. The correlation is
found to be quite stable: for example, if we remove the forces excluded by
the depicted test constraint, then $r=0.990$. The figure also shows the
points calculated with the neutron densities deduced from experiment in
Refs.\ \cite{hof80,zen10,klo07}.}\end{figure}

While one first thinks of using a PREX extraction of $r_n$ to constrain
$\drnp$ of $^{208}$Pb, we show in Fig.~\ref{alrrnp} that $\apv$
and $\drnp$ have themselves a very high linear dependence (the
correlation coefficient is 0.995). The small fluctuation of $\apv$ with
the charge density is more effectively removed by analyzing $\apv$ vs
$r_n-r_p$. Actually, the correlation of $\apv$ and $\drnp$ is
implicit in the BA. That is, expanding Eq.\ (\ref{apvpwba}) at $q\to0$
yields $F_n(q)/F_p(q) \to 1 - (r_n+r_p) (r_n-r_p) q^2/6$, which is driven
by $r_n-r_p$ ($r_n+r_p\simeq 11.1$ fm changes by less than 3\% in the
models). Though Coulomb distortions correct $\apv$ by more than
30--40\%, the correlation prevails in the DWBA result. One sees in
Fig.~\ref{alrrnp} that any nuclear model accurately calibrated to masses
and charge radii nearly falls on the best-fit line and that the confidence
band of the regression is very narrow. Looking at Fig.~\ref{alrrn}, it can
be realized that different models, similarly successful for the well-known
observables, can give the same $\apv$ with different $r_n$
(cf.\ MSkA, BCP, and SkM*; Sk-Rs, Ska, and FSUGold; SkI5 and G2), 
but almost the same $\drnp$ are obtained with these forces. That
the prediction band of the regression is wider horizontally in
Fig.~\ref{alrrn} than in Fig.~\ref{alrrnp} points to the same fact. Thus,
one expects more accurate estimates of neutron observables using the
correlation of Fig.~\ref{alrrnp}. Having found $\drnp$, one can
get $r_n$ by unfolding the finite size of the proton charge from the
accurate \pbx charge radius \cite{ang04}. We note that our
analysis allows one to deduce $\drnp$ and $r_n$ from $\apv$
without assuming any particular shape for the nucleon density profiles.
Altogether, we believe our results firmly back the commissioning of an
improved PREX run where $\apv$ can be measured more accurately. The
present method will permit to retain in $\drnp$ and $r_n$ most of
the experiment's accuracy. As recently proposed \cite{ban10}, if $r_n$ is
first precisely known, then a second measurement can be made at higher
energy to constrain the surface thickness of the neutron density of $^{208}$Pb. 

The correlation of $\apv$ with $\drnp$ is universal in the realm
of mean field theory as it is based on widely different nuclear
functionals. It is of interest to get further indications on it by looking
at existing experiments. The \pbx neutron densities found via proton
elastic scattering at 0.8 GeV in \cite{hof80} and 0.3 GeV in \cite{zen10}
were both deduced from the data in a way consistent with the experimental
charge density of \pbx (known by electron elastic scattering). We
computed $\apv$ using the neutron and charge densities quoted in these
works and plotted the results in Fig.~\ref{alrrnp} against the central
$\drnp$ value of each experiment (0.14 fm in \cite{hof80} and 0.21
fm in \cite{zen10}). We did the same with the data deduced from the
antiprotonic \pbx atom \cite{klo07} (now using the Fermi nucleon
densities of Table VI of \cite{klo07}). It is seen that the theoretical
correlation of the models nicely agrees with these points. Our {\em test}
value $\apv=0.715$ ppm {\em of\/ {\rm 3}\% accuracy} from PREX would
give $\drnp$ as $0.195 \pm 0.057$ fm (see Fig.\ \ref{alrrnp}). As
reviewed in \cite{cen10}, we may recall that the recent constraints from
strong probes, isospin diffusion, and pygmy dipole resonances favor a
range 0.15--0.22 fm for the {\em central} value of $\drnp(
^{208}{\rm Pb})$. Recent informations on the nuclear equation of state derived
from observed masses and radii of neutron stars suggest a similar range
0.14--0.20 fm \cite{stei10,hebe10}.

\begin{figure}[t]
\includegraphics[width=0.95\columnwidth,clip=true]{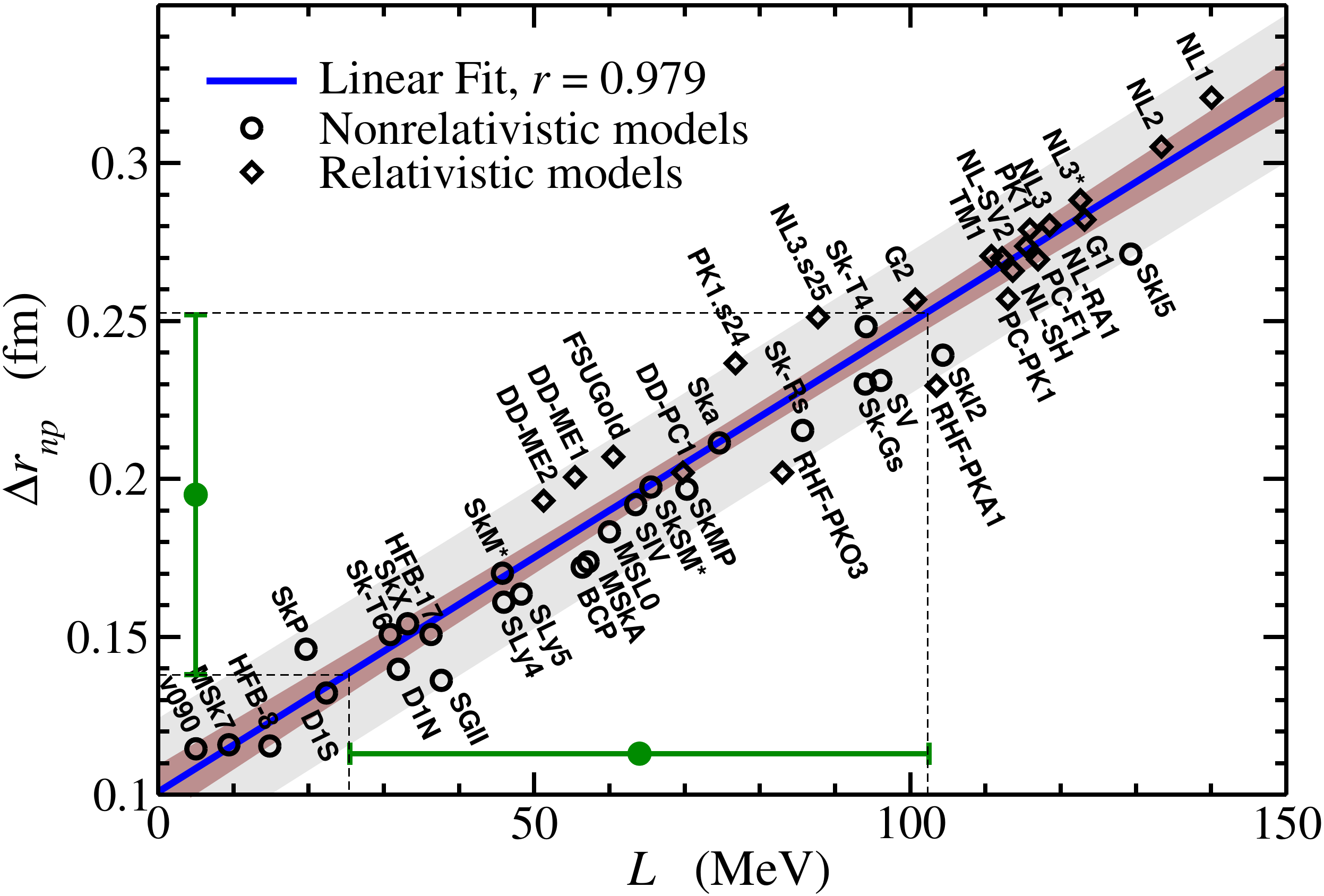}
\caption{\label{alrL}
Neutron skin of \pbx against slope of the symmetry energy. The
linear fit is $\drnp= 0.101 + 0.00147 L$. A sample test constraint
from a 3\% accuracy in $\apv$ is drawn.}\end{figure}

Finally, we analyze how PREX can constrain the density dependence of the
nuclear symmetry energy $E_{\rm sym}(\rho)$ around normal density
$\rho_0$, which is characterized by the slope coefficient $L = 3\rho_0 \,
\partial E_{\rm sym}(\rho) / \partial\rho |_{\rho_0}$ in the literature
\cite{cen09,chen10,carbone10,li08,tsa09}. A larger $L$ value implies a
higher pressure in neutron matter and a thicker neutron skin in
$^{208}$Pb. Interest in $L$ permeates many areas of active research,
such as the structure and the reactions of neutron-rich nuclei
\cite{cen09,todd05,chen10,carbone10,li08,tsa09,ste05a}, the physics of
neutron stars \cite{hor01,xu09,stei10}, and events like giant flares
\cite{flares09} and gravitational radiation from neutron stars
\cite{wen09}. The available empirical estimates span a rather loose range
$30 \lesssim L \lesssim 110$ MeV, with the recent constraints seemingly
agreeing on a value around $L\sim60$ MeV with $\pm$25 MeV spread
\cite{cen09,chen10,carbone10,li08,tsa09}. A microscopic calculation with
realistic nucleon-nucleon potentials and three-body forces predicts $L=66.5$
MeV \cite{vid09}. Figure~\ref{alrL} displays the correlation between
$\drnp( ^{208}{\rm Pb})$ and $L$ \cite{cen09,chen10,carbone10} in the
present analysis. Imposing the previous constraint $\drnp=0.195
\pm 0.057$ fm yields $L=64\pm 39$ MeV. While the central value depends on
our test assumption $\apv=0.715$ ppm, the spread following from a
determination of $\apv$ to 3\% accuracy, essentially does not. Then, we
have to conclude that a 3\% accuracy in $\apv$ sets modest constraints
on $L$, implying that some of the expectations that this measurement will
constrain $L$ precisely may have to be revised to some extent. To narrow
down $L$, though demanding more experimental effort, a $\sim$1\% measurement
of $\apv$ should be sought ultimately in PREX. Our approach can support
it to yield a new accuracy near $\delta\drnp\sim 0.02$ fm and
$\delta L\sim 10$ MeV well below any previous constraint. Moreover,
PREX is unique in that the central value of $\drnp$ and $L$
follows from a probe largely free of strong force uncertainties.

In summary, PREX ought to be instrumental to pave the way for electroweak
studies of neutron densities in heavy nuclei \cite{prex1,prex2,ban10}. To
accurately extract the neutron radius and skin of \pbx from the
experiment requires a precise connection between the parity-violating
asymmetry $\apv$ and these properties. We investigated parity-violating
electron scattering in nuclear models constrained by available laboratory
data to support this extraction without specific assumptions on the shape
of the nucleon densities. We demonstrated a linear correlation, universal
in the mean field framework, between $\apv$ and $\drnp$ that has
very small scatter. Because of its high quality, it will not spoil the
experimental accuracy even in improved measurements of $\apv$. With a
1\% measurement of $\apv$ it can allow to constrain the slope $L$ of the
symmetry energy to near a novel 10~MeV level. A mostly model-independent
determination of $\drnp$ of \pbx and $L$ should have enduring 
impact on a variety of fields, including atomic parity nonconservation 
and low-energy tests of the Standard Model \cite{prex2,vret00,sil05}. 

We thank G. Col\`o, A. Polls, P. Schuck, and E. Vives for valuable
discussions, H. Liang for the densities of the RHF-PK and PC-PK models,
and K. Kumar for information on PREX kinematics. Work supported by the
Consolider Ingenio 2010 Programme CPAN CSD2007 00042 and Grants No.\
FIS2008-01661 from MEC and FEDER, No.\ 2009SGR-1289 from Generalitat de
Catalunya, and No.\ N N202 231137 from Polish MNiSW.


\begin{thebibliography}{00}
\bibitem{enam08} Proc.\ 5th Intl.\ Conf.\ on Exotic Nuclei and
Atomic Masses ENAM'08, Eur.\ Phys.\ J. {\bf A42}, 299 (2009).
\bibitem{hof80} G. W. Hoffmann {\it et al},
Phys.\ Rev.\ {\bf C21}, 1488 (1980).
\bibitem{zen10} J. Zenihiro {\it et al}, 
Phys.\ Rev.\ {\bf C82}, 044611 (2010).
\bibitem{kra04} A. Krasznahorkay {\it et al},
                Nucl.\ Phys.\ {\bf A731}, 224 (2004).
\bibitem{klo07} B. K{\l}os {\it et al},
                Phys.\ Rev.\ {\bf C76}, 014311 (2007).
 (arXiv:0702016)
\bibitem{fried09} E. Friedman, Hyperfine Interact.\ {\bf 193}, 33 (2009). 
 (arXiv:0810.1848)
\bibitem{don89} T. W. Donnelly, J. Dubach, and Ingo Sick, Nucl.\
Phys.\ {\bf A503}, 589 (1989)
\bibitem{vret00} D. Vretenar {\it et al},
                 Phys.\ Rev.\ {\bf C61}, 064307 (2000).
 (arXiv:nucl-th/9911024)
\bibitem{prex2} C. J. Horowitz, S. J. Pollock, P. A. Souder, and
                R. Michaels, Phys.\ Rev.\ {\bf C63}, 025501 (2001).
 (arXiv:nucl-th/9912038)
\bibitem{prex1} K. Kumar, P. A. Souder, R. Michaels, and G. M. Urciuoli, 
                spokespersons, http://hallaweb.jlab.org/parity/prex
             (see section `Status and Plans' for latest updates).
\bibitem{cen10} M. Centelles, X. Roca-Maza, X. Vi\~nas, and M. Warda,
Phys.\ Rev.\ {\bf C82}, 054314 (2010).
 (arXiv:1010.5396)
\bibitem{ang04} I. Angeli, At.\ Data Nucl.\ Data Tables {\bf 87}, 185 (2004)
\bibitem{bro00} B. A. Brown, Phys.\ Rev.\ Lett.\ {\bf 85}, 5296 (2000);
S. Typel and B. A. Brown, Phys.\ Rev.\ {\bf C64}, 027302 (2001).
\bibitem{fur02}  R. J. Furnstahl, Nucl.\ Phys.\ {\bf A706}, 85 (2002).
 (arXiv:nucl-th/0112085)
\bibitem{ste05a} A. W. Steiner, M. Prakash, J. M. Lattimer, and P. J. Ellis, 
                 Phys.\ Rep.\ {\bf 411}, 325 (2005).
 (arXiv:nucl-th/0410066)
\bibitem{todd05} B. G. Todd-Rutel and J. Piekarewicz,
                 Phys.\ Rev.\ Lett.\ {\bf 95}, 122501 (2005).
 (arXiv:nucl-th/0504034)
\bibitem{cen09} M. Centelles, X. Roca-Maza, X. Vi\~nas, and M. Warda,
Phys.\ Rev.\ Lett.\ {\bf 102}, 122502 (2009)
 (arXiv:0806.2886);
M. Warda, X. Vi\~nas, X. Roca-Maza, and M. Centelles,
Phys.\ Rev.\ {\bf C80}, 024316 (2009).
 (arXiv:0906.0932)
\bibitem{carbone10} A. Carbone {\it et al},
                    Phys.\ Rev.\ {\bf C81}, 041301(R) (2010).
 (arXiv:1003.3580)
\bibitem{chen10} L. W. Chen {\it et al},
                 Phys.\ Rev.\ {\bf C82}, 024321 (2010).
 (arXiv:1004.4672)
\bibitem{li08}  B. A. Li, L. W. Chen, and C. M. Ko,
                Phys. Rep.\ {\bf 464}, 113 (2008).
 (arXiv:0804.3580)
\bibitem{tsa09} M. B. Tsang {\it et al},
                Phys.\ Rev.\ Lett.\ {\bf 102}, 122701 (2009).
 (arXiv:0811.3107)
\bibitem{hor01} C. J. Horowitz and J. Piekarewicz,
                Phys.\ Rev.\ Lett.\ {\bf 86}, 5647 (2001).
 (arXiv:astro-ph/0010227)
\bibitem{xu09}  J. Xu et al, Astrophys.\ J. {\bf 697}, 1549 (2009).
 (arXiv:0901.2309)
\bibitem{stei10} A. W. Steiner, J. M. Lattimer, and E.F. Brown,
                 Astrophys.\ J. {\bf 722}, 33 (2010).
 (arXiv:1005.0811)
\bibitem{moya10} O. Moreno, E. Moya de Guerra, P. Sarriguren, and
J. M. Ud{\'\i}as, J. Phys.\ G {\bf 37}, 064019 (2010).
\bibitem{ban10} S. Ban, C.J. Horowitz, and R. Michaels, arXiv:1010.3246. 
\bibitem{regres} N. R. Draper and H. Smith, {\em Applied Regression 
                 Analysis} 3rd.\ ed.\ (Wiley, New York, 1998).
\bibitem{hebe10} K. Hebeler, J.M. Lattimer, C.J. Pethick, and A. Schwenk,
                 Phys.\ Rev.\ Lett.\ {\bf 105}, 161102 (2010).
 (arXiv:1007.1746)
\bibitem{flares09} A.W. Steiner and A.L. Watts,
                   Phys.\ Rev.\ Lett.\ {\bf 103}, 181101 (2009).
 (arXiv:0902.1683)
\bibitem{wen09} D. H. Wen, B. A. Li, and P. G. Krastev,
 Phys.\ Rev.\ {\bf C80}, 025801 (2009).
 (arXiv:0902.4702)
\bibitem{vid09} I. Vida\~na, C. Provid\^encia, A. Polls, and A. Rios,
                Phys.\ Rev.\ {\bf C80}, 045806 (2009)
 (arXiv:0907.1165)
\bibitem{sil05} T. Sil {\it et al}, Phys.\ Rev.\ {\bf C71}, 045502 (2005).
 (arXiv:nucl-th/0501014)
\end{thebibliography}
\end{document}